 \documentclass[final,5p,times,twocolumn,authoryear]{elsarticle}

\usepackage{lipsum}
\usepackage{gensymb}

\usepackage{epsf,palatino,url,wrapfig,array,setspace,amsmath,amssymb,fancyhdr,multirow,lscape,appendix,rotating}
\usepackage{xcolor}
\usepackage{graphicx}
\usepackage{amsmath}
\usepackage{amssymb}
\usepackage{txfonts}
\usepackage{graphicx}
\usepackage{caption}
\usepackage{float}
\usepackage{gensymb}
\usepackage{subfig}
\usepackage{yfonts}
\usepackage{mathrsfs}
\usepackage{newtxtext,newtxmath}
\usepackage{widetext}
\usepackage{flushend}
\usepackage{cuted}
\journal{High Energy Astrophysics}

\begin{document}

\begin{frontmatter}



\title{Quasi-Periodic Oscillations due to radiative feedback mechanism between the disc and corona}


\author[first]{Akash Garg}
\ead{akashgarg.jmi@gmail.com}
\affiliation[first]{organization={Inter-University Center for Astronomy and Astrophysics},
            addressline={Ganeshkhind}, 
            city={Pune},
            postcode={411007}, 
            state={Maharashtra},
            country={India}}
            
\author[first]{Ranjeev Misra}
\ead{rmisra@iucaa.in}


\author[second]{Somasri Sen}
\ead{ssen@jmi.ac.in}

\affiliation[second]{organization={Department of Physics, Jamia Millia Islamia,},
            addressline={Jamia Nagar}, 
            city={New Delhi},
            postcode={110025}, 
            country={India}}
            
\cortext[cor1]{Corresponding author: Akash Garg}            

\begin{abstract}
Compact object systems exhibit Quasi-Periodic Oscillations (QPOs) as revealed by peaked features in their power density spectra. It has been known that stochastic variations in the accretion disc will propagate to the corona after a time delay and that the hard X-rays from the corona impinge back on the disc, giving reflection spectral features. Here, we show that the combination of these two effects makes a simple radiative feedback system between the corona and the disc, which naturally produces the observed QPOs whose primary frequency corresponds to the inverse of the time delay. The analytical form of the expected power spectra can be statistically compared with the observed ones. Hence for the first time, a physical model is used to describe and fit the AstroSat observed power spectra of the black hole systems MAXI J1535-571 and GRS 1915+105, including the QPO, its harmonics as well as the broadband components. 
\end{abstract}




\end{frontmatter}




\section{Introduction}
\label{introduction}

Low-mass X-ray binaries (XRBs) represent stellar interacting binary systems where a compact star, such as a black hole (BH) or a neutron star (NS), and a companion star orbit each other under the effect of their mutual gravitational attractions. Throughout their evolution, these systems transfer gaseous matter from the companion star to the compact star, achieved through either Roche lobe overflow or stellar wind accretion. However, the transferred material must shed a significant amount of its angular momentum before reaching the compact star. This is achieved through an accretion process that begins with the transfer of matter in the form of an elliptical ring, which then undergoes redistribution into a planar disk. As the matter is accreted down the gravitational potential well of the compact star, it loses its energy through viscous dissipation and emits X-ray photons near the compact star \citep{frank22}.

XRBs exhibit rapid variability as revealed by their power spectra, which is the absolute square of the Fourier transforms of their lightcurves \citep{klis89}. The power spectra show complex features such as broadband components often accompanied by peaks which are known as Quasi-Periodic Oscillations (QPOs). The broadband noise is believed to be due to accretion rate fluctuations, which originate in the accretion disc and propagate to the main energy release region \citep{1997MNRAS.292..679L,gilfanov10}.

The QPOs have a well-studied nomenclature defined using properties like centroid frequency ($f_o$), full width at half maximum (FWHM), and fractional RMS normalization. Typically, in BH XRBs, there can be either low frequency (LF) QPOs with $f_o \leq 30$Hz or high frequency (HF) QPOs with $f_o \geq 60$Hz \citep[e.g.][]{2016AN....337..398M,ingram19}. Using the power and width of the observed features in power spectra, LFQPOs are further classified into three types, namely, Type A, B, and C \citep[e.g.][]{wijnand99,casella05}. In NS XRBs, there are three classes of low Hz, Hecto Hz, and kilo Hz QPOs \cite{klis89b}. The frequencies of the QPO should correspond to some intrinsic and characteristic time-scale of the inner accretion disc close to the compact object, and hence, understanding the phenomenon holds the promise of revealing the dynamics of matter in the General relativistic strong gravity regime (\citealp[see for e.g.][and references therein]{ingram19,marco22}).

For example, and for the particular case of LF QPOs in black hole systems, the characteristic frequency may be the Lens-Thirring precession frequency, where the corona in the shape of an inner hot region wobbles in and out of the orbital plane, an effect which happens only in General Relativity for a spinning black hole \citep{1999ApJ...524L..63S,1999ApL&C..38...57S,2014MNRAS.444.2065I}. Alternatively, the QPO frequency may be identified with the inverse of the dynamical time-scale (i.e., the sound crossing time) across the hot inner flow, with General Relativistic corrections \citep{2020ApJ...889L..36M,2021ApJ...909...63L,rawat23}. Correlations of QPO frequencies with low-frequency breaks \citep{wijnand99}, with peaks of the broad components \citep{psaltis99}, with spectral parameters like photon index \citep[e.g.][]{vig03,2019MNRAS.488..720B,2022MNRAS.514.3285G}, and with more source properties \citep[see][for a comprehensive list]{mariano24} have also been used to identify them with characteristic frequencies of the system. 

Apart from frequency identification, the other issue is the dynamic origin of the oscillation. The relatively large amplitudes of these oscillations ($\sim$ few percent) suggest some global mode that encompasses a significant part of the X-ray emitting region.  The QPOs may be due to dynamical oscillation of the corona (or hot inner flow), and attempts have been made based on hydrodynamics to identify global unstable modes \citep{1999ApJ...518L..95T,2004ApJ...612..988T,2007ApJ...663..445S}. Other models such as the Accretion-Ejection Instability \citep{2002A&A...387..497V,1999A&A...349.1003T}, corrugation modes \citep{kato80,kato01}, oscillatory shock model \citep{molteni96,chakra08}, transition layer model \citep{titar04}, pressure modes \citep{cabanac10} and more \citep[see the review][]{belloni16} have also been invoked. However, these models are complicated with uncertainties about whether the assumptions made are valid and whether these instabilities will indeed lead to oscillatory behavior. Moreover, they do not make quantitative predictions regarding the observed power spectra. The Lens-Thirring model of a precessing inner flow \citep{2014MNRAS.444.2065I} is a simpler one but does not provide direct information regarding the amplitude and width of the QPO and its harmonic, which would depend on the driving mechanism and dampening of the precession. The model, however, can be verified by detailed phase-resolved spectroscopy, which for rapid oscillations is challenging.

X-ray spectral analysis of X-ray binaries has, in general, revealed the presence of a disc producing soft photons and a hot corona region, which Comptonizes the disc photons to high-energy ones. While the geometry of the system may change with spectral states (and different geometries have been invoked for the same state), there is near consensus on the presence of a disc and a hot corona \citep[e.g.][]{zdz04,ibragi05,tomsick09,plant15,dbarbara16,kalemci22}. Since X-ray binaries exhibit variability in a wide range of time scales, their origin may be in different regions of the extended disc, and these perturbations propagate to the inner regions \citep{1997MNRAS.292..679L,kotov01,ingram13}. The observed log-normal flux distributions and the flux-rms relationship \citep{uttley01} indicate that such propagating fluctuations do occur in these systems \citep{2005MNRAS.359..345U,heil12,uttley23}. In the inner regions, these fluctuations in the disc are propagated to the corona after a time delay, resulting in long-term variability in hard X-rays. Such a propagating model where variability in the disc induces changes in the corona after a time delay has been used to explain the energy-dependent fractional root mean square (rms) and time-delay in the X-ray emission of X-ray binaries \citep{mubashir16,2019MNRAS.486.2964M,2020MNRAS.498.2757G,jithesh21,2022MNRAS.514.3285G,husain23}.

There is also strong evidence that a significant fraction of the hard X-rays produced in the corona impinge back into the disc, resulting in reflection features such as a broad Iron line and Compton bump \citep{lightman88,fabian89,2007MNRAS.381.1697R,gilfanov10}. The impinging corona flux will be reprocessed by the disc and will enhance the input photon flux into the corona. This effect has been used to explain soft thermal reverberation lags associated with aperiodic variability \citep[e.g.][]{uttley11,kara19,wang21} and the $\sim 50$ microsecond time-lag observed in kHz QPOs of neutron star systems \citep{2001ApJ...549L.229L,2014MNRAS.445.2818K, 2021MNRAS.503.5522K}. This has been extended to QPOs at a few Hz of black hole binaries \citep{garcia21,bellavita22,rawat23b} by invoking large coronal sizes greater than 200 gravitational radii. 

These attempts have been limited to explaining the time lags and not the origin of the QPO. 
\cite{2014MNRAS.445.2818K} showed that the model does not have any natural resonance. In particular, for any assumed oscillation of the coronal heating rate, there is no enhanced flux variation at any frequency, as seen in Figure 3 of \cite{2014MNRAS.445.2818K}. However,  feedback systems, in general, do exhibit resonances, and hence, it is interesting and important to look for modifications of the model that may lead to enhanced oscillations at a particular frequency. \cite{2022A&A...662A.118M} extended the model by considering a finite radiative cooling time-scale of the corona.  They introduced the time-evolution of the temperature of the corona in response to changes in the heating and radiative cooling rate.  They showed that with this modification, the system for some parameter values can exhibit a strong oscillatory behavior.  However, the corona radiative cooling time scale for typical black hole systems is $<$ 0.1 milliseconds. They identify this time-scale with QPO at a few Hz by invoking large coronal size and using accretion rates which would correspond to a luminosity of $10^{35}$ ergs/s, several orders of magnitudes lower than the typical observed values of $10^{38}$ ergs/s.

In this work, we consider a different disc-corona feedback system consisting of two aspects. First, accretion rate variation in the inner disc leads to variation in the coronal heating rate after a time delay. Secondly, the varying coronal radiation impinging on the disc leads to a variation in the local disc accretion rate. If the above two effects are considered together, then the picture arises that fluctuations in the disc are propagated to the corona after a time delay, and subsequently, the variation in the corona affects the disc, forming a disc-corona feedback system. In the next section, we develop an analytical model to describe such a system and in Section 3, we fit its prediction to the observed power spectra of black hole binaries and discuss the implications of the results in Section 4.

\begin{figure}
 \centering
  \includegraphics[width=0.48\textwidth]{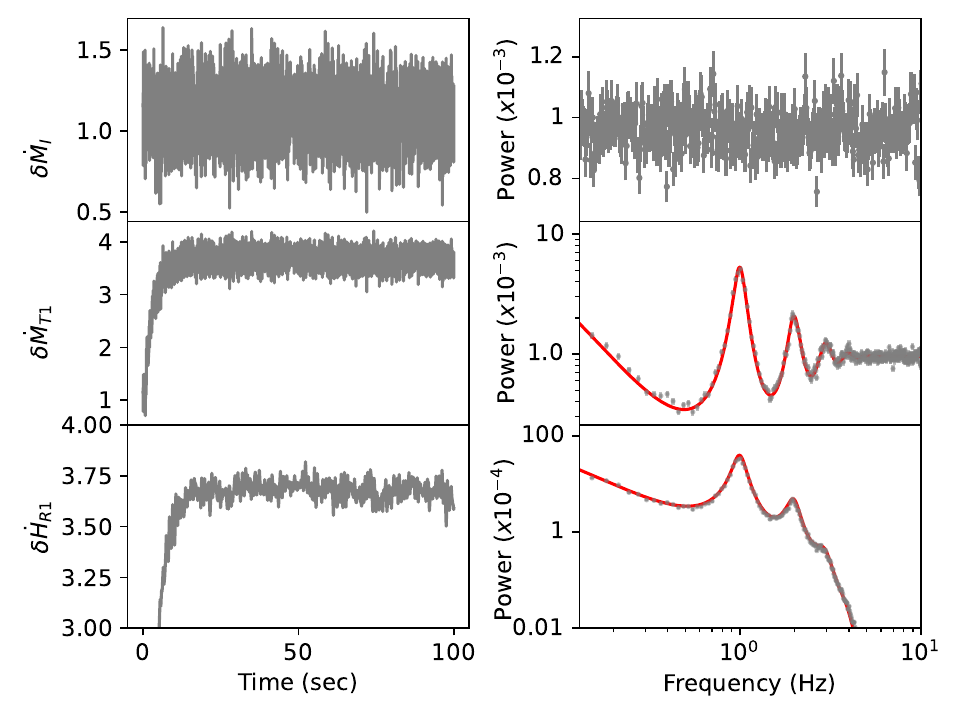}
  \caption{Top left panel shows the first 100 seconds of simulated time series for intrinsic white noise variation of the local accretion rate $\delta \dot M_{I} (t)$ and the corresponding power spectrum is shown in the top right panel. The middle and bottom left panels show the resulting responses in accretion rate variation $\delta \dot M_{T1} (t)$ and heating rate variation $\delta \dot H_{R1} (t)$. The simulated power spectra shown in the middle and bottom right panels are compared with those obtained from the square of the analytical forms of $\delta \dot {\mathscr{M}_{T1}} (\omega)$ and $ \delta \dot {\mathscr{H}_{R1}} (\omega)$ respectively represented by Equations \ref{mdot}.  }
  \label{simul}
\end{figure}

\section{Disc-Corona Feedback System}
We adopt the physical picture where there is a truncated accretion disc, and in the inner region, there is a hot corona. Variations of the local accretion rate $\dot M$ at the inner disc radius, $R_{in}$, would then induce a change in the heating rate of the corona $\dot H$ after a time delay. In particular, an accretion rate variation normalized to its average value, $\delta \dot M \equiv \Delta \dot M / <\delta \dot M>$ will produce a linear response in the normalized heating rate variation such that

\begin{equation}
\delta \dot H_R (t) = A \int^{\infty}_{0} G_A(t,t^\prime)\delta \dot M(t^\prime) dt^\prime
\label{eqn1}
\end{equation}

where $G_A$ is the response function, $\delta \dot H_R (t)$ is the response in the coronal heating rate variation, and A is the normalisation factor. For simplicity, we assume the response to be a Gaussian function, i.e. $G_A(t,t^\prime) =\frac{1}{\sqrt{2\pi} \sigma_A} e^{-(t-t^\prime-\tau_A)^2/2\sigma_A^2}$ (where the peak is centered at $\tau_A$ and $\sigma_A$ represents the standard deviation (or width) of the Gaussian) and discuss the implications of this assumption later. This implies that a spike (i.e. a delta function) variability in $\dot M(t^\prime)$ would produce a Gaussian evolution of $\delta \dot H_R (t)$ after a time delay $\tau_A$ and that in the limit $\sigma_A \rightarrow 0$, the relation reduces to $\delta \dot H_R(t) = A \delta \dot M(t-\tau_A)$. In Fourier space\footnote{We have used different styling of all variables in frequency space.}, this will be,
$ \delta \dot {\mathscr{H}_R} (\omega) = A \mathscr{G}_A \delta  \dot {\mathscr{M}} (\omega)$, where
$\mathscr{G}_A =  e^{-i\omega\tau_A} e^{-(\omega \sigma_A)^2/2}$.

We further consider that the variation in the heating rate produces a change in the hard X-ray spectra impinging back onto the inner disc, resulting in a response in the local accretion rate such that $\delta \dot {\mathscr{M}_R} (\omega) = B \mathscr{G}_B \delta \dot {\mathscr{H}_R} (\omega)$, where $ \mathscr{G}_B = e^{-i\omega\tau_B} e^{-(\omega \sigma_B)^2/2}$ (the peak is centered at $\tau_B$ and $\sigma_B$ represents the standard deviation (or width) of the Gaussian) and B is the normalisation factor. The varying coronal radiation will impinge back on the inner truncated disc, which may lead to changes in the local accretion rate. The irradiation may affect the local thermal profile in radius or height, which in turn may change the local accretion rate. The effect of X-ray irradiation on the outer ($>$ 500 $r_g$) regions of the disc has been studied (e.g., \cite{cunning}, \cite{bari15}). It is shown by \cite{bari15} that a varying inner disc X-ray flux can lead to local accretion rate variation in the outer disc, which in turn can propagate to the inner regions. However, this occurs in the long viscous time scales of the outer disc. The hard X-ray emission from the corona may then perhaps also influence the local accretion rate in the inner truncated disc regions. However, a physical model to verify this process is complex and would require a detailed MHD simulation, which also incorporates radiative transfer in the disc. Nevertheless, it is possible that the local accretion rate in the truncated disc varies in response to variations in the irradiating flux from the corona. We also note that the thermal response of the disc to rapid variations in coronal output is commonly invoked to describe the observed X-ray reverberation lags in BH XRBs (e.g., Uttley et al. 2011; Kara et al. 2019; Wang et al. 2021; De Marco et al. 2022). Although a quantitative estimate of the effect would require detailed hydrodynamic simulations, here we have parameterized the effect using the form above, $B \mathscr{G}_B$. If there is an intrinsic stochastic variation in the accretion rate $\delta \dot {\mathscr{M}_I}$, then the total variation would be $\delta \dot {\mathscr{M}_T} = \delta \dot {\mathscr{M}_I} + \delta \dot {\mathscr{M}_R}$, where $ \delta \dot {\mathscr{M}_R}$ is the response i.e.  $\delta \dot {\mathscr{M}_R} = B \mathscr{G}_B \delta \dot {\mathscr{H}_R} (\omega) = AB \mathscr{G}_A \mathscr{G}_B \delta \dot {\mathscr{M}_T} (\omega)$. This yields

\begin{eqnarray}
  \delta \dot {\mathscr{M}_{T1}}   =  \frac{\delta \dot {\mathscr{M}_{I}}}{1-ABe^{-i\omega (\tau_A+\tau_B)} e^{-(\omega (\sigma_A+\sigma_B))^2/2}} \nonumber \\
  \delta \dot {\mathscr{H}_{R1}}   =   \frac{A e^{-i\omega\tau_A} e^{-(\omega \sigma_A)^2/2} \delta \dot {\mathscr{M}_{I}}}{1-ABe^{-i\omega (\tau_A+\tau_B)}e^{-(\omega (\sigma_A+\sigma_B))^2/2}}
  \label{mdot}
\end{eqnarray}

\begin{figure}
\centering
  \includegraphics[width=0.48\textwidth]{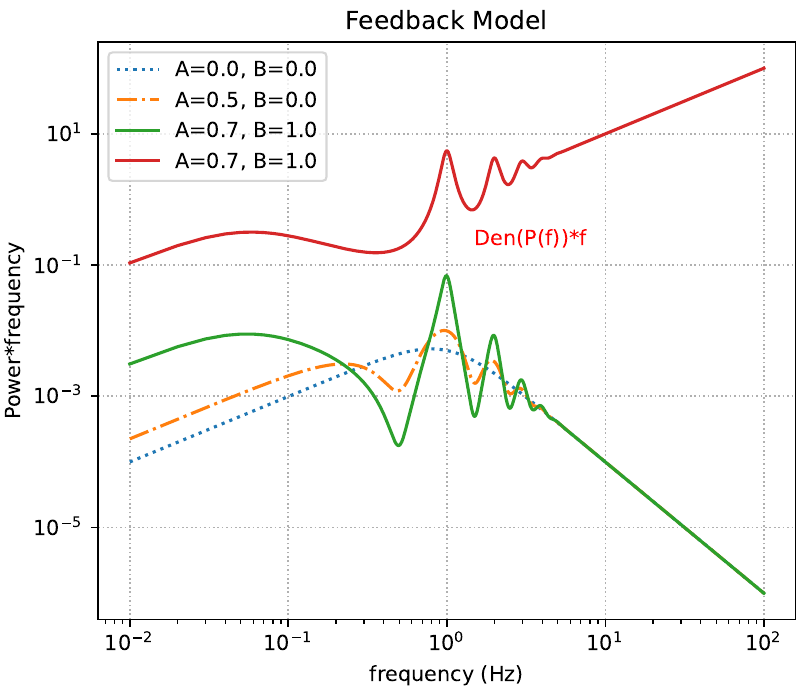}
  \caption{The expected power spectra of a system with fiduciary parameters illustrate the emergence of QPOs due to disc-corona feedback. The radiation flux is taken to be  $\delta \mathscr{F} = \delta \dot {\mathscr{H}} + \delta  \dot {\mathscr{M}}$ and only intrinsic variation of the local accretion rate is considered of the form $|\delta \dot {\mathscr{M}_{I}}|^2 = 0.01/(1+(f)^{p_1})$. If there was only this intrinsic variation, then the expected power spectrum would be a broad component, as shown by the dotted line. The dependence of the coronal heating rate on the accretion rate after a time delay of 1 second would result in a power spectrum with broad features at $\sim$ 1 Hz and harmonics, as shown by the dashed-dotted line. The additional feedback dependence of the accretion rate on the coronal heating gives narrow, coherent QPO features, as shown by the solid green line. The solid red line shows the denominator of Equation \ref{finaleqn} and illustrates the importance of the parameter 'B' in the feedback case.}
  \label{model}
\end{figure}

\begin{figure*}
\centering
  \includegraphics[width=0.25\textwidth,height=5cm]{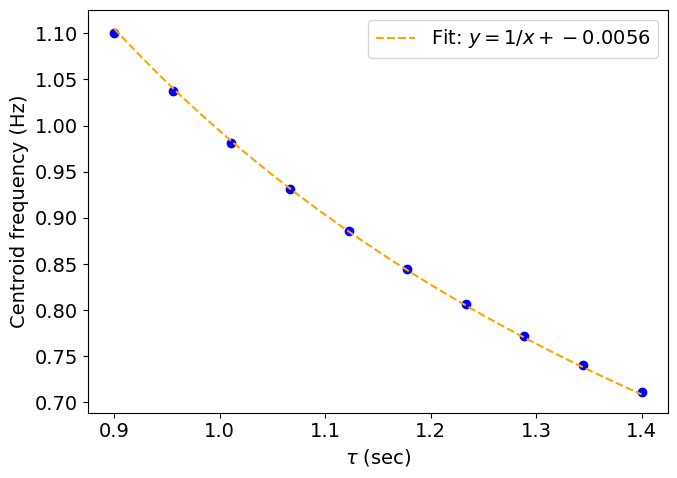}
  \includegraphics[width=0.35\textwidth,height=5cm]{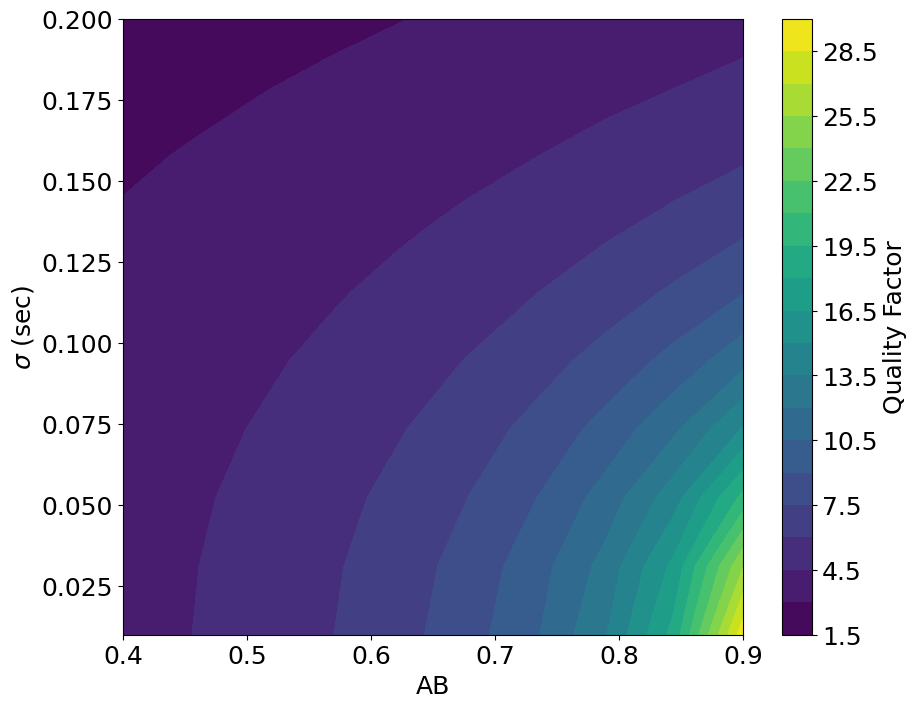}
  \includegraphics[width=0.35\textwidth,height=5cm]{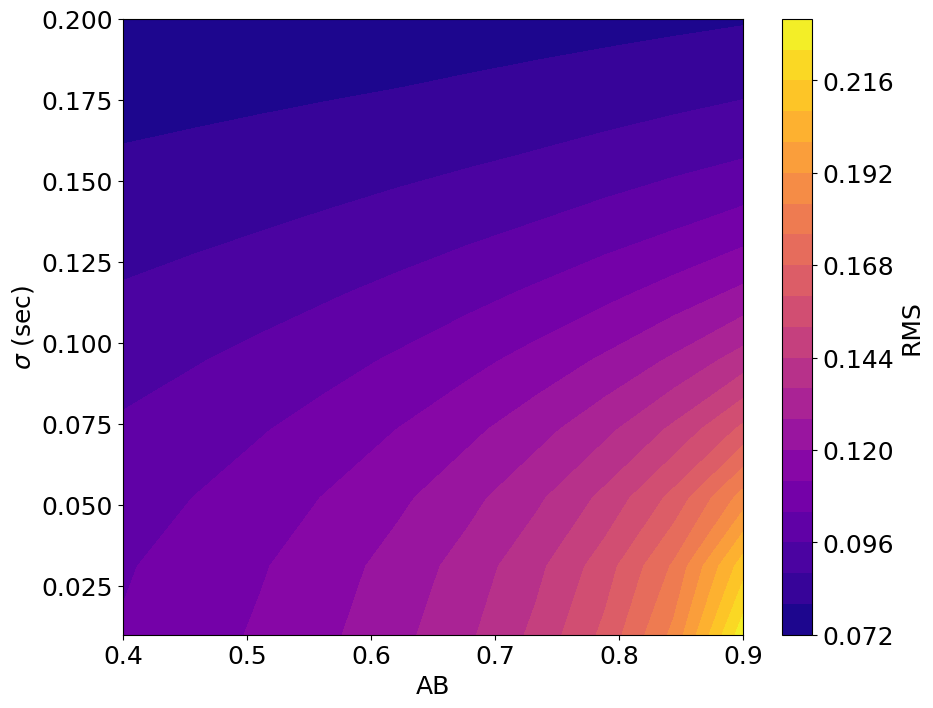}
  \caption{The left panel shows the correlation between the centroid frequency of QPO ($E_L$) and the feedback model parameter $\tau$. There is a clear inverse relation between the two parameters, which can be fitted using the inverse function, y=1/x-0.056. The middle and right panels show the contour plots between the two feedback model parameters $\sigma$ and AB for different values of Quality factors and rms, respectively.}
  \label{corr}
\end{figure*}

where the subscript $1$ denotes that these variations are due to the intrinsic variation in accretion rate $\delta \dot {\mathscr{M}_{I}}$. To validate the above equations and to illustrate the phenomenon, we simulated a white noise time series to represent the intrinsic stochastic variation $\delta \dot M_{I} (t)$. Then we computed $\delta \dot H_{R1} (t)$, using Equation \ref{eqn1} (with $A=1$, $\tau_A = 1$ and $\sigma_A = 0.1$) and considered the response of the accretion rate to the heating rate to be instantaneous, i.e.  $\delta \dot M_{R1} (t) = \delta \dot H_{R1} (t)$. The left panel of Figure \ref{simul} shows the first 100 seconds of the simulation, while the right panel shows the corresponding power spectra for the full-time series. As expected, the power spectra obtained from the simulated time series are exactly identical to the square amplitudes of the functions given in Equation \ref{mdot}. Note that the variability of the responses $\delta \dot H_{R1} (t)$ and  $\delta \dot M_{R1} (t)$ are quasi-periodic in the sense that the behavior is seen as harmonic peaks in the power spectra and not apparent in the time series themselves.

Analogously any intrinsic variation in the corona heating rate $\delta \dot {\mathscr{H}_{I}}$, would result in\\

\begin{eqnarray}
  \delta \dot {\mathscr{H}_{T2}} = \frac{\delta \dot {\mathscr{H}_{I}}}{1-ABe^{-i\omega (\tau_A+\tau_B)} e^{-(\omega (\sigma_A+\sigma_B))^2/2}}
\end{eqnarray}

\begin{eqnarray}
     \delta \dot {\mathscr{M}_{R2}} = \frac{B e^{-i\omega\tau_B} e^{-(\omega \sigma_B)^2/2}\delta \dot {\mathscr{H}_{I}}}{1-ABe^{-i\omega (\tau_A+\tau_B)}e^{-(\omega (\sigma_A+\sigma_B))^2/2}}
\end{eqnarray}

where the subscript $2$ denotes that these variations are due to the intrinsic variation in corona heating rate $\delta \dot {\mathscr{H}_{I}}$.
The observed flux, $F$ in some energy band, is assumed to have a linear response to the accretion rate and the coronal heating rate, i.e., $\delta \mathscr{F} = C\delta \dot {\mathscr{H}} +D \delta  \dot {\mathscr{M}}$, where C and D are normalisation factors. Thus, the power spectrum $P(f) = |\delta \mathscr{F}|^2$ from such a system can be written as

\begin{widetext}
\begin{equation}
\scalebox{0.9}{$
\begin{aligned}
P(f) = \Bigg(&\frac{(AC/D)^2 e^{-\omega^2\sigma_A^2}+ 2(AC/D)\cos(\omega\tau_A)e^{-\omega^2\sigma_A^2/2}+1}
{1 + (AB)^2e^{-(\omega (\sigma_A+\sigma_B))^2}-2 AB\cos(\omega (\tau_A+\tau_B))e^{-(\omega (\sigma_A+\sigma_B))^2/2}}\Bigg)D^2\delta \dot{\mathscr{M}}^2_I \\
+ \Bigg(&\frac{(BD/C)^2 e^{-\omega^2\sigma_B^2}+ 2(BD/C)\cos(\omega\tau_B)e^{-\omega^2\sigma_B^2/2}+1}
{1 + (AB)^2e^{-(\omega (\sigma_A+\sigma_B))^2}-2 AB\cos(\omega (\tau_A+\tau_B))e^{-(\omega (\sigma_A+\sigma_B))^2/2}}\Bigg)C^2\delta \dot{\mathscr{H}}^2_I
\end{aligned}
$}
\label{finaleqn}
\end{equation}
\end{widetext}

The intrinsic variations may be described as broadband noise with a break frequency i.e. $|\delta \dot {\mathscr{M}_{I}}|^2 = N_1/(1+(f/f_1)^{p_1})$ and $|\delta \dot {\mathscr{H}_{I}}|^2 = N_2/(1+(f/f_2)^{p_2})$, where $f_1$ and $f_2$ are the break frequencies.

Figure \ref{model} shows the expected power spectrum from such a disc-corona feedback system with only intrinsic variation in the accretion rate for some fiduciary values of $f_1 = 1.0$ Hz, $p_1 = 3.0$, $N_1 = 0.01$, $C = D = 1$, $\tau_A = 1.0$, $\tau_B = 0.0$, $\sigma_A = 0.1$ and $\sigma_B = 0$. The dotted line shows the resultant power spectrum when the accretion and heating rates are not related ($A = B = 0$), showing just the broad component. The dashed-dotted line shows the case when the heating rate depends on the accretion one, i.e., $A = 0.5$, while there is no feedback $B = 0$. Broad features at $\sim 1 $ Hz appear due to the cosine term in the numerator of Equation \ref{finaleqn}. Finally, the solid line shows the spectrum when $AB = 0.7$, and one sees the appearance of narrow coherent QPO and its harmonics due to the form of the denominator in Equation \ref{finaleqn}.

We further explore the parameter space of the feedback model to understand the impact of its parameters, such as AB, $\tau = \tau_A+\tau_B$, and $\sigma = \sigma_A+\sigma_B$, on the observed properties of the QPOs like Centroid frequency, Quality factor, and the fractional variability (root mean square (rms) value). We begin by simulating the power spectra of the feedback model for various parameter values and then fit the simulated QPO feature in the spectra using a Lorentzian line profile. The Lorentzian line profile includes fitting parameters such as $F_l$, $width$ (FWHM line width in Hz), and a normalization (norm). Subsequently, we analyze the correlation between the feedback model parameters and $F_l$ (centroid frequency), Quality factor ($=F_l$/$width$), and rms ($=\sqrt{norm}$) of the predicted QPO.

\begin{figure*}
\centering
  \includegraphics[width=0.48\textwidth]{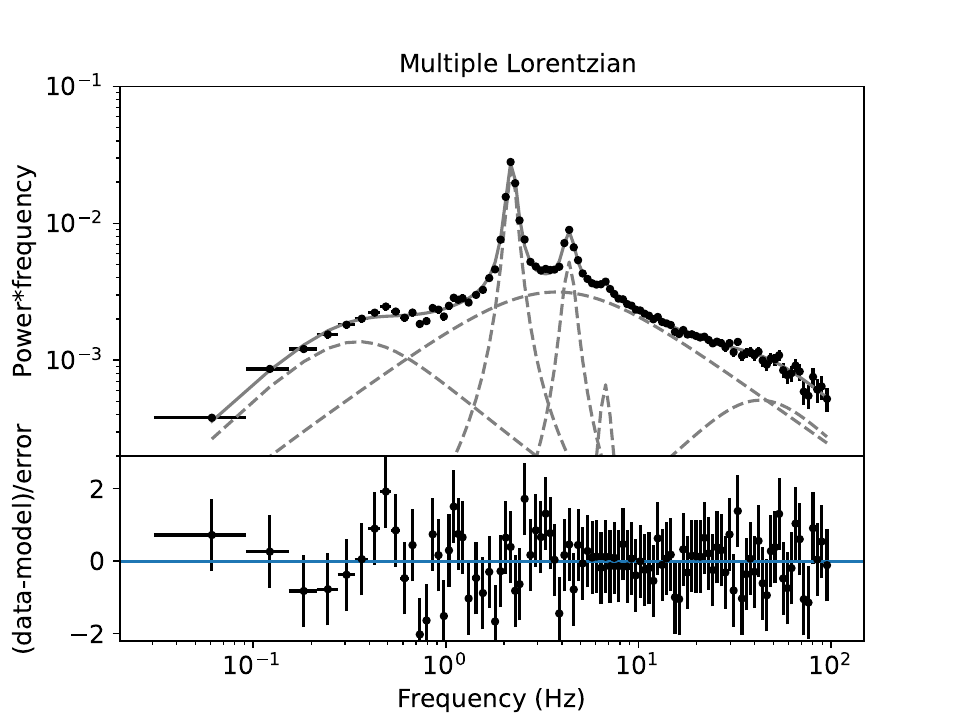}
  \includegraphics[width=0.48\textwidth]{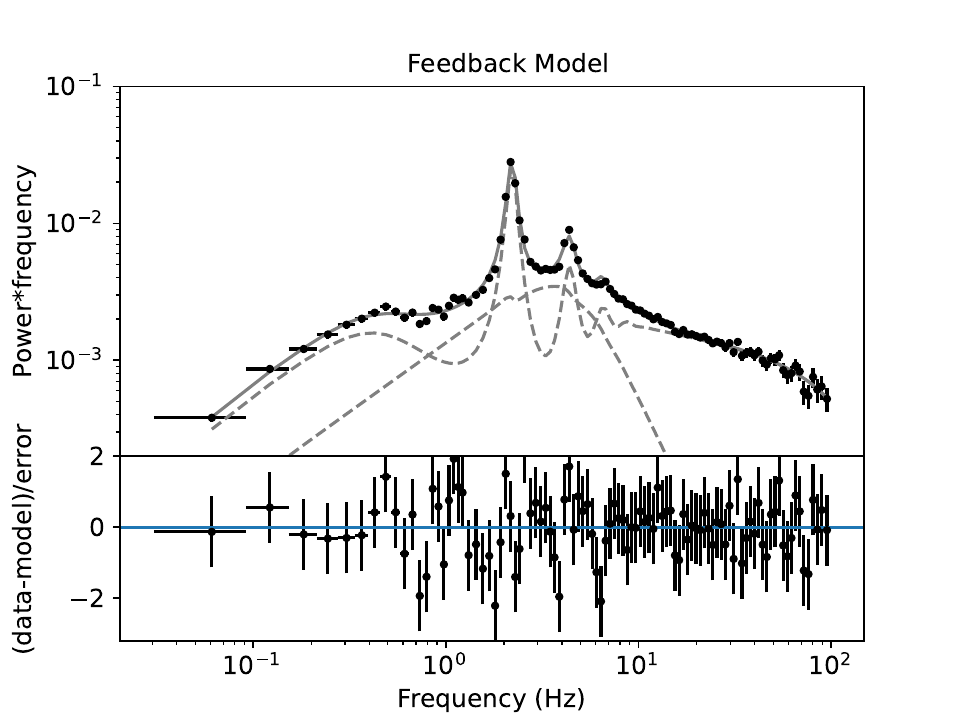}
  \caption{The observed power spectrum for MAXI J1535-571 in the energy range 4-20 keV using AstroSat/LAXPC observation is plotted in both panels (black-filled circles). In the left panel, the spectrum is fitted using a phenomenological model consisting of six Lorentzian components having 17 parameters, which gives a $\chi^2/dof$ of $55/83$. In contrast, the right panel shows the same spectrum fitted using the disc-corona feedback model having only 12 parameters and gives a $\chi^2/dof$ of $71.8/88$. The parameter values with errors are listed in Table 1. }
  \label{maxi}
\end{figure*}

Figure \ref{corr} shows the relation between the feedback model parameters and QPO properties. We find, as expected, that the centroid QPO frequency is $ F_l= 1/\tau = 1/(\tau_A+\tau_B)$. The Quality factor Q depends on the multiplication of the responses, i.e., $AB$, and the sum of the dampening factors $\sigma = \sigma_A + \sigma_B$. The middle and right panels of Figure \ref{corr} display contour plots, color-coded to represent varying values of the Quality factor and rms. The primary result here is that a QPO, as defined as having Q $>$ 3, is expected when $AB \gtrsim 0.5$, but very large (Q$>$20) can also be obtained for $AB \sim 1$ and small values of $\sigma$.  With these correlations, we provide the necessary conditions for a QPO to be observed due to feedback, which may facilitate any future interpretations.


\section{Comparison with Observations}

\begin{figure*}
\centering
  \includegraphics[width=0.48\textwidth]{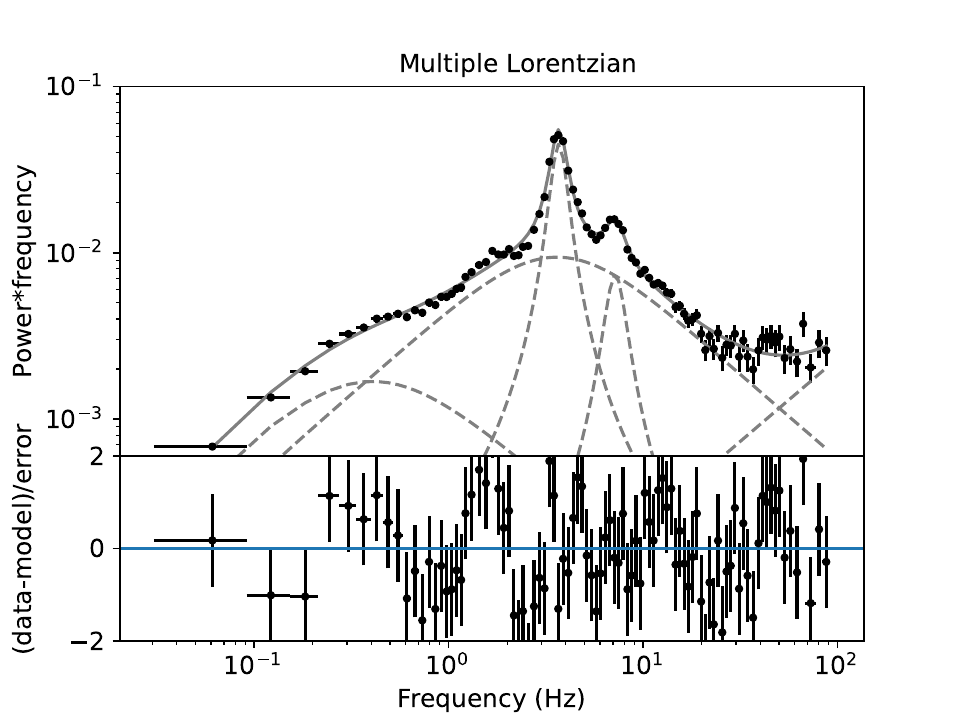}
  \includegraphics[width=0.48\textwidth]{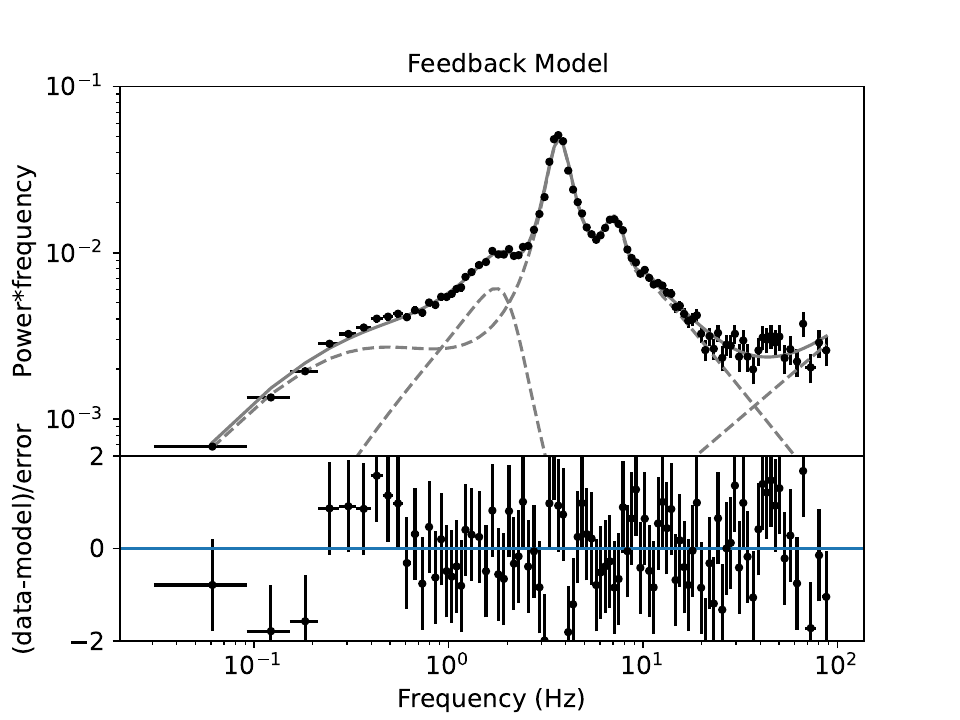}
  \caption{The observed power spectrum for GRS 1915+105 in the energy range 4-20 keV using AstroSat/LAXPC observation is plotted in both panels (black-filled circles). In the left panel, the spectrum is fitted using a phenomenological model consisting of four Lorentzian components having 12 parameters. In contrast, the right panel shows the same spectrum fitted using the disc-corona feedback model, having the same number of parameters but providing a significantly better fit with a $\chi^2/dof$ of $77/81$ as compared to $108/82$. The parameter values with errors are listed in Table 2. A constant (power-law component with index zero) has been included in both fittings to take into account uncertainties in the dead-time corrected Poisson noise level.}
  \label{grs}
\end{figure*} 

In this section, we compare the predicted power spectrum from a disc-corona system using AstroSat observations. AstroSat is an Indian multiwavelength observatory equipped with five scientific instruments: a Soft X-ray telescope (SXT), three Large X-ray proportional counters (LAXPC), a cadmium-zinc-telluride imager (CZTI), and two Ultraviolet imaging telescopes (UVIT), along with a scanning sky monitor (SSM). The first four instruments are co-aligned to observe celestial sources across a broad X-ray energy range of 0.3-80 keV, simultaneously capturing near- and far-UV wavelengths \citep{singh14,singh22}. The three LAXPC detectors are nominally identical and provide a good collective area, a modest energy resolution, and a high timing resolution of 10 $\micro$s in the 3-80 keV band, which makes the detectors suitable to do the timing studies \citep{2016SPIE.9905E..1DY,2017JApA...38...30A,antia17,antia21}.

Over the past 8 years, LAXPC observed multiple X-ray binary systems and detected the presence of strong low-frequency Type C QPOs \citep[][in GRS 1915+105]{yadav16,2019ApJ...870....4R,2020MNRAS.498.2757G,rawat22,athulya22}, \citep[][in MAXI J1535-571]{2019MNRAS.487..928S,2019MNRAS.488..720B,2022MNRAS.514.3285G,chatt21}, \citep[][in Swift J1658.2-4242]{jithesh19,bogen20}, \citep[][in H 1743-322]{chand21,husain23}, \citep[][in GRS 1716-249]{rout21,chatt21a}, \citep[][in MAXI J1348-630]{jithesh21}, \citep[][in MAXI J1803-298]{chand22,jana22}. In this work, we choose to fit the LAXPC PDS of two black hole systems, MAXI J1535-571 and GRS 1915+105, which contain Type C QPOs along with harmonics and broadband noise. We chose both sources as examples to check the validity of the feedback model for all features in PDS in a broad frequency band.

AstroSat/LAXPC observed MAXI J1535-571 during its 2017 outburst from September 12-17, 2017, through the Target of Opportunity proposal (Observation ID: T01\_191T01\_9000001536). The source was detected in the hard-intermediate state, and the power spectrum analysis revealed the presence of strong QPOs with harmonics in the frequency range $\sim$ 1.7-3.0 Hz \citep{2019MNRAS.487..928S,2019MNRAS.488..720B}. \cite{2019MNRAS.488..720B} analyzed the whole set of observations by dividing them into 66 segments with similar exposures and found a sharp QPO in each of the segments \citep[see Table 1 in ][]{2019MNRAS.488..720B}. For our purpose, we select one of the 66 segments and extract the PDS for further analysis. AstroSat/LAXPC observed GRS 1915+105 at multiple instances in different spectral states. We consider a particular observation of GRS 1915+105 taken by AstroSat on March 28 2017 (Observation ID: 20170328\_G06\_033T01\_9000001116). \cite{2019ApJ...870....4R,yadav16} detected the source in $\chi$ class \citep[see][for the existence of 14 different classes of lightcurves for GRS 1915+105]{belloni20} and found the presence of narrow QPOs in the frequency range $\sim$ 3.3-3.7 Hz in the LAXPC PDS.

For both the observations of MAXI J1535-571 and GRS 1915+105, we extract the PDS for the chosen segments using the subroutine {\sc{laxpc\_find\_freqlag}} of the LAXPC software\footnote{\url{http://Astrosat-ssc.iucaa.in/laxpcData}}. We choose a minimum frequency resolution of 0.05 Hz and generate the PDS for both sources in a frequency band of 0.05-100 Hz and an energy range of 4-20 keV. As observed before, we find a QPO at $\sim$ 2.18 Hz along with two more harmonics and broadband noise and a QPO at $\sim 3.66$ Hz along with a harmonic and broadband noise in LAXPC PDS of MAXI J1535-571 and GRS 1915+105 respectively as shown in Figures \ref{maxi} and \ref{grs}.

We begin by fitting the PDS of MAXI J1535-571. Traditionally, the empirical method to quantify such a PDS is to use a number of Lorentzian components. For this particular data set, even after including a systematic uncertainty of 5\%, six Lorentzian components (with 17 free parameters) are required to fit the shape of the PDS, yielding a chi-square of 55 for 83 degrees of freedom (Left panel of Figure \ref{maxi}). 

Furthermore, since we have an analytical Equation~\ref{finaleqn} that predicts the power spectra, we directly defined a model based on this equation using the \texttt{mdefine} command in the {\sc{XSPEC}}\footnote{\url{https://heasarc.gsfc.nasa.gov/docs/xanadu/xspec/}} package. This provided us with an option to use our defined model as other {\sc{XSPEC}} models. We fitted the data with this new model, which has 12 parameters, and found a good description of the observed PDS of MAXI J1535-571, giving a chi-square of 71.8 for 88 degrees of freedom (Right panel of Figure \ref{maxi}). For the Lorentzian fit, if one of the three broad components is removed, the chi-square becomes 188 for 86 degrees of freedom. The best-fit parameters with errors are shown in Table 1. 

Next, we fit the PDS of GRS 1915+105. For this data set, a phenomenological fit with four Lorentzians results in a chi-square of 108 for 82 degrees of freedom with 12 parameters (Left panel of Figure \ref{grs}). A better fit is obtained when the feedback model is used, resulting in a chi-square of 77 for 81 degrees of freedom (Right panel of Figure \ref{grs}) with the best-fit parameters listed in Table 2.

\begin{table*}
	\centering
	\caption{Best fit model parameters for both models for MAXI J1535-571}
	\label{tab:spectramaxi}
	\begin{tabular}{c c c c c c} 
		\hline \hline
		Lorentzian & Parameter & Value & Feedback Model Parameter & Value \\
		\hline
		L1 & Centroid freq. (Hz)    &  $2.18^{+0.01}_{-0.01}$                 & AB                          & $0.867^{+0.001}_{-0.004}$\\
		   & Width (Hz)             &  $0.26^{+0.03}_{-0.02}$                 & AC/D                        & $-0.867^{+0.002}_{-0.003}$\\
		   & Norm (x $10^{-3}$)     &  $4.92^{+0.34}_{-0.32}$                 & $\tau_A$ (ms)                    & $449.7^{+2.0}_{-1.9}$\\
     \\
		L2 & Centroid freq. (Hz)    &  $4.34^{+0.04}_{-0.04}$                 & $\sigma_A$  (ms)              & $50.82^{+1.98}_{-1.98}$\\
     & Width (Hz)             &  $0.65^{+0.15}_{-0.12}$                 & $f_1$ (Hz)                  & $4.47^{+0.22}_{-0.21}$ \\
     & Norm (x $10^{-3}$)     &  $1.26^{+0.19}_{-0.17}$                    & $p_1$                       & $3.98^{+0.41}_{-0.33}$\\
\\
     L3 & Centroid freq. (Hz)    &  $6.69^{+0.25}_{-0.36}$                 & $N_1$ (x $10^{-3})$                       & $1.3^{+0.1}_{-0.1}$ \\
     & Width (Hz)             &  $0.85^{+1.60}_{-0.62}$                         & $\tau_B$ (ms)                    & $2.1^{+1.1}_{-0.3}$ \\
     & Norm (x $10^{-4}$)     &  $1.38^{+1.7}_{-0.81}$                    & $\sigma_B $ (ms) & $2.36^{+1.54}_{-0.75}$ \\   
     \\
	    L4 & Centroid freq. (Hz)    &  $0.0$ (f)                              & $f_2$ (Hz)                  & $0.47^{+0.03}_{-0.03}$ \\
     & Width (Hz)             &  $7.49^{+0.71}_{-0.61}$                 & $p_2$                       & $3.28^{+0.03}_{-0.03}$ \\
     & Norm (x $10^{-3}$)     &  $9.90^{+0.71}_{-0.83}$                    & $N_2$                       & $207.17^{+51.34}_{-114.81}$ \\   		   
\\
  L5 & Centroid freq. (Hz)    &  $25.86^{+9.12}_{-18.41}$ \\  
     & Width (Hz)             &  $>53.64$ \\	
     & Norm (x $10^{-3}$)     &  $1.12^{+0.41}_{-0.24}$ \\	
\\	
  L6 & Centroid freq. (Hz)    &  $0.15^{+0.04}_{-0.06}$ \\
     & Width (Hz)             &  $0.62^{+0.12}_{-0.10}$ \\
     & Norm (x $10^{-3}$)     &  $3.47^{+0.51}_{-0.43}$ \\		   
\\
  $\chi^2$/dof  &             &  55/83 = 0.66 &                                                & 71.81/88 = 0.82\\
		\hline
	\end{tabular}
    \\Note that (f) indicates the parameter was frozen.
\end{table*}

\begin{table*}
	\centering
	\caption{Best fit model parameters for both Models for GRS 1915+105}
	\label{tab:spectragrs}
	\begin{tabular}{c c c c c c} 
		\hline \hline
		Lorentzian & Parameter & Value & Feedback Model Parameter & Value \\
		\hline
		L1 & Centroid freq. (Hz) &  $3.66^{+0.02}_{-0.02}$         &    AB           & $0.52^{+0.03}_{-0.03}$\\
		   & Width (Hz)  &  $0.77^{+0.08}_{-0.07}$                 &    AC/D         & $-1.27^{+0.56}_{-0.28}$\\
     & Norm   &  $0.014^{+0.001}_{-0.001}$                   &    $\tau_A$ (ms)     & $191.4^{+19.4}_{-12.9}$\\
\\	
  L2 & Centroid freq. (Hz) &  $7.12^{+0.12}_{-0.13}$         &    $\sigma_A$ (ms) & $< 14.7$\\
     & Width (Hz) &  $1.89^{+0.59}_{-0.46}$                  &    $f_1$ (Hz)   & $2.002^{+0.044}_{-0.063}$ \\
     & Norm (x $10^{-3}$)   &  $2.9^{+0.8}_{-0.7}$           &    $p_1$        & $9.78^{+5.53}_{-2.09}$\\
\\
     L3 & Centroid freq. (Hz) &  $0.56^{+0.71}$                 &    $N_1$ (x $10^{-3}$)        & $2.45^{+0.54}_{-0.47}$ \\
     & Width (Hz) &  $7.12^{+0.40}_{-0.51}$                  &    $\tau_B$ (ms)    & $77.71^{+8.67}_{-28.12}$ \\
     & Norm (x $10^{-3}$)  &  $27.7^{+1.4}_{-1.4}$           &    $\sigma_B$ (ms) & $37.89^{+2.91}_{-3.04}$ \\   
\\
  L4 & Centroid freq. (Hz) &  $0.0$ (f)                      &    $f_2$ (Hz)   & $3.92^{+1.18}_{-0.03}$ \\
 & Width (Hz) &  $0.81^{+0.17}_{-0.16}$                  &    $p_2$        & $2.47^{+0.26}_{-0.24}$ \\
     & Norm   &  $0.005^{+0.001}_{-0.001}$                   &    $N_2$ (x $10^{-3}$)      & $8.23^{+7.20}_{-2.61}$\\
     \\
		Powerlaw & Norm (x $10^{-5}$) &  $2.32^{+0.45}_{-0.45}$    &      & $3.23^{+0.60}_{-0.65}$ \\
\\	
  $\chi^2$/dof & & 108/82 = 1.31 &                           & 77/81 = 0.95\\
		\hline
	\end{tabular}
    \\Note that (f) indicates the parameter was frozen.
\end{table*}

For both the data sets, the time delay between the accretion rate in the disc and the heating rate of the corona, $\tau_A$ is of the order of seconds and significantly larger than the time delay between the coronal heating rate and the disc accretion rate, $\tau_B$ which is in the order of milliseconds. Thus, the frequency of the QPO is mostly determined by the former, i.e. $\sim 1/\tau_A$. The QPO frequency has been identified as the inverse of the sound crossing time $\sim R_{in}/c_s$, where $R_{in}$ is the truncated inner disc radius and $c_s$ the sound speed at that radius \citep{2020ApJ...889L..36M}. Thus, in this interpretation, a disturbance caused by the accretion rate variation in the disc travels at the speed of sound to the corona, resulting in a heating rate variation. On the other hand, the disc accretion rate reacts on a much faster time scale of milliseconds to the coronal heating variation, $\tau_B,$ indicating that this could be a combination of the light travel time and the response of the disc. Indeed, if this feedback is due to the irradiation of the disc by the corona, this is the time scale that one may expect. The parameters $AB$ and $AC/D$ are dimensionless quantities and, hence, as expected, are of order unity. We note that here, we have employed a specific interpretation that a feedback between the disc accretion rate and the coronal heating is driving the QPO. There may be other parameters, such as inner disc radius and the optical depth, which may also vary with time delays, leading to a more complex scenario.

\section{Discussion}

The disc-corona feedback system described in this work provides a natural framework to explain the observed Quasi-periodic Oscillations (along with harmonics) in X-ray binary systems. Despite its simplicity, the model provides an analytical form for the shape of the power spectrum, which can be quantitatively tested against data. Two examples of fitting the power spectra for two black hole systems have been shown in this work. These fittings are statistically better or equal to fits undertaken using empirical Lorentzian components.

The values of the parameters obtained from the fitting and their variation for different observations, in principle, can be identified with physical processes to validate the model. For instance, the parameter $\tau_A$, which is the time taken for a disc perturbation to reach the corona, may be associated with a physical time scale of the system. Indeed, it has been shown that the QPO frequency (which is approximately $1/\tau_A$ in this framework) may be identified as the sound crossing time at the truncated inner disc radii \citep{2020ApJ...889L..36M}. Moreover, the hard X-ray lags between the disc and coronal variability in BH XRBs \citep[e.g.][]{uttley11,marco15} can be associated with our parameter $\tau_A$. Such hard lags are often linked to the diffusion timescales over which the accretion rate fluctuations can propagate to the inner region of the accretion flow around the compact object \citep{nowak96,1997MNRAS.292..679L,kotov01}. The assumed geometry is that of a truncated disc with a hot corona inside. Local stochastic accretion rate variation in the inner regions of the disc will propagate to the corona, causing a variation in the heating rate of the corona, like in the stochastic propagation model proposed by \cite{1997MNRAS.292..679L}. In \cite{1997MNRAS.292..679L}, this propagation is due to viscous effects, and hence, the time-scale of the propagation is the local viscous time-scales. However, it may also be possible for these disturbances in the inner disc to propagate on shorter sound travel time-scales.

The parameter $\tau_B$ is the time delay between the coronal variation and the consequent effect on the disc. This may be a combination of the light travel time of the coronal radiation to reach the disc and the thermal time scale of the disc to adjust to the incident radiation. Note that for MAXI J1535-571, the best fit $\tau_B$ turns out to be 2 msec, which, if it is only due to light travel time, would imply a distance of 600 km or 20 Gravitational radii for a ten solar mass black hole. However, for GRS 1915+105, the best fit $\tau_B$ is significantly larger $\sim$ 78 msecs, which would imply large distances that may not be physical. Thus, the time delay may include other effects, such as the thermal time-scale of the disc for the accretion rate to change. As mentioned earlier, the time-scale estimation would require a detailed hydrodynamical simulation.

We would like to emphasize that this current work represents an initial effort to use propagating variations to describe the observed QPOs and their harmonics in the power spectra. As an instance, we have checked the validity of the model for the limiting number of cases of the observed power spectra of GRS 1915+105 and MAXI J1535-571, and it may or may not perform well for all kinds of oscillations as observed during the outburst. However, future studies building on this idea will further test its consistency and potentially establish more robust insights. More extensive analysis of different observations of a source along with photon spectral parameters (such as the optical depth of the corona, inner disc radii, accretion rate, etc.) can be undertaken to see if the values of $\tau_B$ are physical or not. Perhaps a more straightforward analysis would be to estimate the parameters $C$ and $D$, which relate to how variations in the accretion rate and heating rate of the corona translate to variation in the observed flux in an energy band, which may be inferred from the photon spectral model (e.g., \cite{2020MNRAS.498.2757G}). Since the parameters used for fitting the power spectra are AB and AC/D, these estimates can be used to infer parameters $A$ and $B$, which relate to the strength of the connection between the disc and the corona. The model has the potential to be used to fit the power spectra for different energy bands and to predict the time lag as a function of frequency for these different energy bands. Thus, more advanced versions of the model would take into account the time-averaged energy spectrum of the system, along with the energy-dependent timing property to identify the physical parameters in the disc and the corona that are coupling with each other to produce the QPO phenomenon. This provides an incentive to develop physically motivated forms for the intrinsic broadband variability of the disc and corona rather than the empirical form used here. Similarly, the response function has been assumed to be a Gaussian, but in reality, it may be more complex, such as the Fast Rise Exponential Decay (FRED) form. The framework described provides important input to hydrodynamic or magneto-hydrodynamic simulations of accretion discs around black hole systems. The results suggest that if radiative feedback from the corona to the disc is taken into account, the simulation may indeed show oscillatory behavior, which can be compared with observations.

\section*{Acknowledgements}
We are grateful to an anonymous reviewer for their constructive comments, which helped us significantly improve the quality of the manuscript. This research work is utilizing ASTROSAT/LAXPC data available at the Indian Space Science Data Centre (ISSDC). The work has made use of software provided by the High Energy Astrophysics Science Archive Research Center (HEASARC).

\bibliographystyle{elsarticle-harv} 
\bibliography{references}






\end{document}